\documentclass[12pt,preprint]{aastex}

\begin{document}

\title{$Y^{2}$ Isochrones with an Improved Core Overshoot
Treatment}

\author{Pierre Demarque \& Jong-Hak Woo}
\affil{Department of Astronomy, Yale 
University, New Haven, CT 06520-8101, USA}

\author{Yong~-Cheol Kim}
\affil{Department of Astronomy, Yonsei University,
Seoul 120-749, Korea}

\and

\author{Sukyoung K. Yi}
\affil{Department of Physics, University of Oxford, Oxford, OX1~3RH, UK}

\begin{abstract}
Convective core overshoot affects stellar evolution rates and the dating 
of stellar populations.
In this paper, we provide a patch to the $Y^2$ isochrones 
with an improved treatment 
of convective core overshoot.  The new tracks cover the transition 
mass range from no convective core to a fully developed convective core.
We compare the improved isochrones to CMDs of a few well observed open star 
clusters in the Galaxy and the Large Magellanic Cloud. Finally we discuss 
future prospects for improving the treatment of core overshoot with the 
help of asteroseismology.
\end{abstract}


\section{Introduction}

Understanding the physics of convective core overshoot is important
in interpreting the color-magnitude diagrams (CMDs) and the
luminosity functions of open star clusters, and in dating young and 
intermediate age stellar populations.  
Core overshoot has several effects on stellar evolution (see e.g. 
Stothers 1991).
Among the most notable are the effects on the shape of the main sequence
turnoff, on the rate of evolution in the main sequence and  
subgiant phases, and on the 
ratio of total lifetimes spent in the core hydrogen burning phase 
and the shell hydrogen burning stage. The implications of convective 
core overshoot for deriving a chronology of open star clusters in the Galaxy 
have long been recognized (Maeder \& Mermilliod 1981).
Detailed comparisons of theoretical isochrones with star cluster CMDs 
suggest the need to include convective overshoot in stellar evolution 
calculations.  Gaps in the stellar distribution
near the main sequence, which are identified with the 
hydrogen exhaustion phase, 
have been observed 
in the CMDs of open star clusters. 
The location and size of these gaps  yield better 
agreement with those theoretical isochrones
that admit some amount of core overshoot (Stothers \& Chin 1991; 
Carraro et al. 1993; Daniel et al. 1994;
Demarque, Sarajedini \& Guo 1994; Kozhurina-Platais et al. 1997;
Nordstr\"{o}m, Andersen \& Andersen 1997).

The purpose of this paper is to improve on the treatment of convective 
core overshoot in the $Y^2$ isochrones (Yi et al. 2001, hereafter Paper I; 
Kim et al. 2003, hereafter Paper II). We focus on the mass range where 
convective cores begin to appear in stellar models near the main sequence.
This transition region affects sensitively the morphology of isochrones and 
the luminosity function near the main sequence turnoff. As a result,
it affects the ages derived for stellar populations based on the CMDs
 of star clusters.
It also modifies the predicted integrated spectral energy distributions 
of stellar populations and their derived spectral ages in 
studies of distant galaxies (Yi et al. 2000). 

In view of the 
importance of core overshoot for intermediate age stellar populations, we have 
calculated a patch to the original $Y^2$ isochrones to prevent the troublesome 
distortions that occurred in some instances 
near the turnoff in the original isochrones. 
Most of the interest in core overshoot in the grids of evolutionary tracks 
available in the literature has concentrated 
on massive stars with well developed convective cores (Maeder \& Meynet 1988;
Bertelli et al. 1990; Stothers \& Chin 1991; Meynet et al. 1994).
Even in the Geneva grids, in which intermediate mass evolutionary tracks have 
been included, no mention is made on the problems posed 
for isochrone construction by the narrow transition 
from the absence to the presence of a convective core as the mass increases
(Schaller et al. 1992; Charbonnel et al. 1993).
In a more recent study based on the Padova approach to core overshoot, 
Girardi et al. (2000) describe a treatment of the transition region 
in which the chosen critical transition mass, taken to be 1.0$M_{\odot}$, 
is independent of chemical composition.  

It is however well-known that the mass at which a convective core first 
appears along the main sequence is 
a sensitive function of chemical composition. This result must be taken into 
account in any detailed comparison with observation.  
Already, the original isochrones of Papers I and II addressed this chemical  
composition dependence by introducing   
a critical  mass $M^{conv}_{crit}$, defined as 
the mass above which stars continue to have a substantial convective core 
even after the pre-MS phase of evolution is ended.
 
In addition to taking into account the dependence of $M^{conv}_{crit}$ on 
chemical composition, our new evolutionary tracks also include
several improvements which are physically more realistic, as
detailed in \S4.  These changes are useful 
for chemical compositions that 
differ markedly from solar and for applications which 
require interpolation between the isochrones.

\S2 summarizes the input physics and parameters used in the previous 
$Y^2$ isochrone papers. In \S3, we discuss the physics of 
convective core overshoot and the procedures used to 
model its structural and evolutionary effects in stars. Semi-empirical 
estimates of the 
extent of core overshoot based principally on the CMDs of star clusters 
are given in \S4. In \S5, we then describe the new 
overshoot approach adopted in the critical mass transition region.
Some comparisons between observed open cluster CMD's are presented in 
\S6.  Finally, future prospects for 
a better physical understanding of convective core overshoot are briefly 
summarized in \S7.

\section{Input physics and parameters of the $Y^2$ isochrones}

The evolutionary tracks presented in this paper were constructed using 
the Yale Stellar Evolution Code YREC with the  
same input physics and parameters as in Paper I  
and Paper II.
A more complete discussion of the choice of parameters can be found in 
these two papers.  A useful assessment of validity of the $Y^2$ stellar 
models and other models including a detailed comparison with 
observational data for stars in the pre-main sequence and early post-main 
sequence phases of evolution   
has recently been published by Hillenbrand \& White (2004).  
 
All evolutionary tracks were 
started at the pre-main sequence birthline. In order to 
preserve internal consistency with the rest of the 
$Y^2$ database, 
the same set of chemical compositions were used. 
The solar mixture of Grevesse, Noels \& Sauval (1996) 
was adopted for 
the heavy elements abundances, which yields a metal-to-hydrogen ratio 
$(Z/X)_{\odot}$ = 0.0244. 
The mixing length parameter 
in the convective 
envelopes, calibrated on the Sun, was also 
kept unchanged, i.e. $\it l/H_{p}$ = 1.7432.    
Consistent with the solar calibration and the 
initial chemical composition $(Y,Z)_{0}$ = (0.23, 0.00),
we adopted the chemical enrichment formula 
$Y = 0.23 + 2Z$.  
Following Paper II, we constructed evolutionary tracks for $\alpha$-enhanced  
mixtures corresponding to [$\alpha$/Fe] = 0.0, 0.3 and 0.6, respectively.
Table~1 lists the chemical compositions of the evolutionary tracks constructed 
for this paper.   
A more complete discussion of the input parameters can be found 
in Papers I and II.

\subsection{Microscopic physics}
In the interior, the OPAL radiative opacities 
(Rogers \& Iglesias 1995; Iglesias \& Rogers 1996) for 
the appropriate mixtures were used.  
At low temperatures, the opacities from Alexander \& Ferguson (1994)
were used for the scaled 
solar mixtures. For the $\alpha$-enhanced mixtures, the low temperature 
opacity tables specifically calculated for Paper II by Alexander 
were adopted.  
The conductive opacities from Hubbard \& Lampe (1969) were adopted 
for log$\rho$ $\leq$ 6.0 and from Canuto (1970) in the relativistic 
regime where log$\rho$$>$ 6.0. 
The OPAL equation of state was taken from the work of 
Rogers, Swenson \& Iglesias (1996).  
Helium diffusion was included in the evolutionary calculations with the 
help of the Bahcall-Loeb formula (Bahcall \& Loeb 1990; 
Thoul, Bahcall \& Loeb 1994).
As before, the energy generation routines of Bahcall \& Pinsonneault 
(1992) and the 
cross-sections listed in Bahcall (1989) have been used.

\subsection{Color transformations}

Theoretical properties ($[Fe/H], L, T_{eff}$) have been transformed into 
colors and magnitudes using the semi-empirical 
color transformation tables of Lejeune, Cuisinier \& Buser 
(1998; hereafter LCB) and of Green, Demarque \& King (1987; hereafter GDK).  
The tables have been normalized so that the absolute visual magnitude of the 
Sun ($M_{v\odot}$) becomes 4.82 (Livingston 2000), regardless of the 
table used.
The color transformation is for the filter systems of 
$(UBV)_{Johnson}$$(RI)_{Cousins}$ in the GDK table and of   
$(UBV)_{Johnson}$$(RI)_{Cousins}$$(JHKLM)_{ESO}$ in the LCB table. 
A fuller discussion can be found in Paper I.

\section{Convective core overshoot}
\subsection{The physical process}

Convective core overshoot is understood here as the presence 
of material motions and/or mixing beyond the canonical 
boundary for convection defined by the classic Schwarzschild 
criterion (1906). 
Early investigations by Roxburgh (1965), who used the mixing length 
theory, and by Saslaw \& Schwarzschild (1965), who based
their argument on thermodynamic grounds (i.e. the edge of the 
convective core in
massive stars is sharply defined in an entropy diagram), suggested
that little overshoot takes place at the edge of convective cores.
But Shaviv \& Salpeter (1973)
pointed out subsequently that if one takes into account the presence of
hydrodynamic motions and turbulence, one might expect some
non-negligible amount of core overshoot. 

More recently, Zahn (1991) has discussed  the 
complex physical interaction between convective and radiative 
transfer in the overshoot region.
It turns out that the nature of the overshoot depends
sensitively on the details of the local physics at the 
convective core edge. The 
local P\'{e}clet number, which
characterizes the relative importance of radiative and turbulent
diffusivity, determines whether the temperature gradient in the 
overshoot region is adiabatic (the situation called 
``penetration'' by Zahn), or whether the local 
radiative transfer dominates the energy
transport. In the latter case, overshoot does 
not modify the stable temperature
gradient beyond the Schwarzschild limit, and its main effect 
is to cause mixing into the radiatively stable layers just 
outside the convective core edge. This 
kind of overshoot has been described as ``overmixing''.

It follows from the above discussion that 
a proper treatment of convective core overshoot requires a 
radiative hydrodynamic treatment near the convective boundary. 
Kuhfu\ss~ (1986) has proposed a non-local treatment 
based on the anelastic approximation for convection and the diffusion 
approximation for radiation. An application of the Kuhfu\ss~ theory 
to describe overshoot from convective stars of the upper 
main sequence has been made by Straka et al. (2004), who 
proposed some seismic tests to test the theory. 

\subsection{Core overshoot in stellar models}  

In the absence of a general theory of convective 
overshoot  which would enable us to calculate the amount of core 
overshoot for a star of a given mass and chemical composition, 
a semi-empirical phenomenological approach must be used 
in calculations of stellar evolution.  
Several computational
schemes of various degrees of sophistication in treating the
physics of overshoot have been discussed in the 
literature (Prather \& Demarque 1974; Maeder 1975; Maeder \&
Meynet 1988; Bressan, Bertelli \& Chiosi 1981; 
Bertelli et al. 1990; Straka et al. 2004). 
The most common parameterization of core overshoot, and the one 
that was adopted in the $Y^2$ evolutionary tracks, is to 
evaluate the overshoot length in terms 
of the  local pressure scale height $H_{p}$ at or across the 
formal edge of the convective core, where the core 
edge is defined by the Schwarzschild (1906) criterion. The amount of 
core overshoot is then given by the product $\Lambda_{OS}H_{p}$,
where $\Lambda_{OS}$ is a constant parameter less than unity. 
In the $Y^2$ models, the temperature 
gradient in the overshoot region was assumed to be unaffected by 
the overshoot, and the main result was an 
extension of the mixed region beyond the edge of the convectively 
unstable core. 

The present paper is primarily concerned with the calculation 
of core overshoot for stars with masses in the vicinity of 
$M^{conv}_{crit}$,
the critical mass at which a convective core makes its 
appearance on the main sequence. This simple parameterization of 
core overshoot in terms of the pressure scale height breaks down 
as the convective core vanishes because $H_{p}$ $\rightarrow$ $\infty$
as $r$ $\rightarrow$ 0 (see e.g. Wuchterl \& Feuchtinger 1998).

Roxburgh's integral constraint is another way to quantify the 
amount of core overshoot since it provides an upper limit to the 
radial extent of convective penetration 
(Roxburgh 1989,1998; Zahn 1991; Canuto 1997).  Because it only provides 
an upper limit, the value of Roxburgh's integral must in practice 
be multiplied 
by an adjustable parameter to achieve good agreement with observations 
of cluster CMDs. This free parameter,  which is analogous to the 
$\Lambda_{OS}$ parameter discussed above, was evaluated for the 
open cluster NGC 6819 to be about 0.5 (Rosvick \& VandenBerg 1998). 

It is often desirable to combine the theoretical limit to core overshoot 
set by Roxburgh's integral constraint and the convenience of the 
pressure scale height approach in stellar evolution calculations.  
Recent examples can be found in the work of Woo \& Demarque (2001) 
and Di Mauro et al. (2003). 
The Woo-Demarque approach, however, requires a much finer stellar 
mass grid of  
evolutionary tracks, than is available for the $Y^2$ isochrones.  This 
is  because the transition in the Woo-Demarque treatment takes place in a 
mass range narrower than the mass grid size available in the 
$Y^2$ database (0.1$M_{\odot}$).

\section{Empirical estimates of the extent of core overshoot}

Observations of star cluster CMDs provide a useful guide in 
evaluating $\Lambda_{OS}$ for a large grid of 
stellar evolutionary tracks.  The semi-empirical approach 
of comparing 
synthetic CMDs based on theoretical isochrones to the CMDs and 
luminosity functions of observed star clusters is adopted.
Studies of star cluster CMDs 
indicate that $\Lambda_{OS}$ $\simeq$ 0.2$H_{p}$ for  
clusters with ages in the range 1-3 Gyr, with turnoff 
masses $M_{TO}$ in the range 1.4-2.0 $M_{\odot}$,
and that OS $\simeq$ 0.0-0.1$H_{p}$ for older clusters (4-6 Gyr and
$M_{TO}$ = 1.1-1.3 $M_{\odot}$)
(Stothers 1991; Demarque, Sarajedini \& Guo 1994; 
Kozhurina-Platais et al. 1997).  This statement is compatible with 
the conclusion discussed in the previous section that if the 
radial extent of core overshoot is expressed in terms of the 
local pressure height at the core edge, the effective 
value of the parameter $\Lambda_{OS}$ 
must decrease as the core radius decreases.

There are other independent ways to determine $\Lambda_{OS}$.  
Detached eclipsing binaries in which the components 
have convective cores are consistent with 
some core overshoot, of the order of 0.2$H_{p}$
for stars with fully developed convective cores 
(Ribas, Jordi, \& Gim\'{e}nez 2000).  
It is important to remember, however, that such overshoot estimates
depend on all other features of the standard stellar models being correct.
There are several uncertainties in stellar models that 
may affect our core overshoot estimates.  The boundaries of 
convective regions in stellar models can be 
sensitive to the adopted 
composition parameters, such as helium abundance, [Fe/H] and 
[$\alpha$/Fe], all of which entail uncertainties.
And even if the composition were perfectly well-known,  
Stothers \& Chin (1991) have pointed out the high sensitivity of 
core overshoot to the adopted opacities. For example,
increases in radiative opacities from the OPAL group (Iglesias \&
Rogers 1996) over the previous generation Los Alamos Opacity Library
(Huebner et al. 1977) decrease the need for overshoot in comparison
with observational data. 

Internal rotation may also play a role in core overshoot.
Deupree (1998, 2000) has considered the combined effects of rotation and
convective overshoot in massive stars, which have large convective cores,
based on the two-dimensional hydrodynamic simulations. Rotation could  
be significant in the case of less massive stars as well (in the range
$1.5-2.0 M_{\odot}$), with shear-driven turbulence near the edge of the
convective core mixing material from the helium-enriched core into
the envelope (e.g. see rotating star models by Pinsonneault et al. 1991).
Rotationally induced mixing could thus in effect 
enlarge the size of the mixed convective core.

By providing a powerful independent way of probing the stellar interior, 
stellar seismology will provide new tools to measure the extent of convective 
regions and overshooting.    
The promise of seismology will be briefly 
discussed in the last section of this paper.

\section{Isochrone construction}

\subsection{Original $Y^2$ treatment}
 
Guided by the observational studies mentioned above, 
the published $Y^2$ isochrones were 
constructed assuming $\Lambda_{OS}$ = 0.2 in stars with well developed 
convective cores.  For convenience, and also based on the 
available observational data, the isochrones
were divided into two classes, i.e. young isochrones
($\leq$ 2 Gyr), in which full core overshoot
was taken into account, and old isochrones ($\geq$ 3 Gyr)
in which no overshoot (i.e. $\Lambda_{OS}$ = 0.0) 
was included regardless of the stellar mass.
By inspection of the no overshoot evolutionary tracks, 
we first found the critical mass $M^{conv}_{crit}$ above which stars 
continue to have a substantial convective core even after the pre-MS 
phase is ended.  The critical mass $M^{conv}_{crit}$ is listed in
Table~1 for each chemical composition.  Evolutionary tracks were 
constructed including core overshoot for stars with masses in excess of 
$M^{conv}_{crit}$.  The young isochrones were then constructed using 
the evolutionary tracks with overshoot for $M$ $\geq$ $M^{conv}_{crit}$, and 
evolutionary tracks with no overshoot for lower masses.   
Note that    
the mass interval between our adjacent models in the grid of 
evolutionary tracks being 0.1 $M_{\odot}$,
the value of $M^{conv}_{crit}$ effectively used in the interpolation 
is between the listed value of $M^{conv}_{crit}$ and 
($M^{conv}_{crit}$ $-$ 0.1).  
For isochrones older than 3 Gyr, we used stellar models 
with no overshoot regardless of mass.  This procedure is satisfactory 
for stars with metallicities close to that of the Sun.  For example,  
for solar composition $M^{conv}_{crit}$ is near 1.2 $M_{\odot}$,
which means that the procedure seems acceptable when $M_{TO}$ $\geq$ 
($M^{conv}_{crit}$ + 0.2). 

Subsequent interpolations of the $Y^2$ isochrones in a variety of 
research applications revealed that 
this step function approach could create 
locally unphysical distortions of the isochrone and luminosity function. 
The assumptions made in the original $Y^2$ isochrones can be invalid 
for chemical compositions very 
different from the solar abundance.
In addition, setting 
the transition age at 3 Gyr, which was selected on the basis of 
the available cluster CMD's, is unrealistic for high metallicity 
isochrones.  The ``transition age'' $t_{trans}$ 
(i.e. the age at which convective cores 
begin to affect the turnoff morphology) increases sharply with 
increasing metallicity.  For Z = 0.0001, $t_{trans}$ is 1.2 Gyr, whereas 
for Z = 0.04, $t_{trans}$ reaches 7 Gyr. 

\subsection {Improved core overshoot treatment}

The improvement in our overshoot treatment is two folds.
First, we take into account the metallicity dependence of 
$M^{conv}_{crit}$. As a result, the transition age $t_{trans}$
defined in \S5.1 is no longer fixed at 3~Gyr but changes as 
a function of metallicity.
Second, we include the condition that the extent of core overshoot 
should decrease to zero as the convective core radius approaches zero.
As noted above, this change is required
for the most metal-rich mixtures, in which small convective cores may
still be present near the turnoff for ages well above 3~Gyr.
The step function has now been replaced by a
smoother function for the overshoot parameter in the evolutionary calculations.
The precise dependence and the numerical procedure are
chosen for compatibility with the rest of the $Y^2$ evolutionary tracks,
within the simple pressure scale height
description of overshoot used in the models.
A grid of evolutionary tracks
has been constructed in the transition mass region, and
a revised set of isochrones was then calculated
which incorporates the improved evolutionary tracks.

For the purpose of this study, 
we have adopted the following prescription for ramping between 
$M^{conv}_{crit}$  and ($M^{conv}_{crit}$ $+$ 0.2):\\\\
$\Lambda_{OS}$ = 0.0~for~~~~~$M$~$<$ $M^{conv}_{crit}$\\
$\Lambda_{OS}$ = 0.05~for~~~~$M$ = $M^{conv}_{crit}$\\
$\Lambda_{OS}$ = 0.1~for~~~~~$M$ = $M^{conv}_{crit}$ + 0.1\\
$\Lambda_{OS}$ = 0.15~for~~~~$M$ = $M^{conv}_{crit}$ + 0.2\\
$\Lambda_{OS}$ = 0.2~for~~~~~$M$~$>$ $M^{conv}_{crit}$ + 0.2\\

The values of $M^{conv}_{crit}$ used in the evolutionary tracks are 
given in Table~1, as function of chemical composition.
They were then combined with the 
existing evolutionary tracks calculated with $\Lambda_{OS}$ = 0.2 for 
$M$~$>$ $M^{conv}_{crit}$ + 0.2, and with $\Lambda_{OS}$ = 0.0 for 
$M$~$<$ $M^{conv}_{crit}$.  
New isochrones were constructed using the same isochrone generation codes 
as in Papers I and II only after minor alterations.

The four panels of Figure~1 compare the transition region for isochrones 
before and after applying the patch, for Z = 0.0001, 0.001, 0.004 and 
0.02, respectively.  
A major usefulness of the overshoot-ramping is that it
allows a much smoother age interpolation between isochrones when the
population's turnoff mass is near $M^{conv}_{crit}$, that is, near the
transition age.
Figure~2 illustrates the improvement. The old and new overshoot
treatments make little or no changes for ages larger than the transition age
because such old isochrones are constructed solely based on the stellar 
models without a convective core.
In the case of Z = 0.001, the transition age is roughly 1.9--2.0~Gyr. 
As expected, little change has been made to the 4~Gyr isochrone. 
The problems in the previous isochrones are 
however clearly visible in the interpolated isochrones for smaller ages.
The two intermediate-age isochrones (1.8 and 2.5~Gyr) were produced 
using the interpolation routine provided in Paper 1.
Three features, marked as (a), (b) and (c), summarize the problems
with the Paper 1 isochrones.
Feature (a) in dotted line shows a second dip below the major
main-sequence dip.
This is simply an artifact caused by the previous delta function 
treatment on the overshoot parameter. Feature (b) clearly shows 
how the interpolation fails for the age near the transition age. 
Lastly, feature (c) shows a failure 
in the temperature interpolation on the red giant branch.
All these problems have now been solved by the introduction of the 
ramping treatment on the overshoot. In the rest of this paper, 
we shall refer to the patched isochrones as $Y^2$ OS isochrones,
which can be foud at any of the three $Y^2$ isochrone websites:\\
http://csaweb.yonsei.ac.kr/~kim/yyiso.html\\ 
http://www-astro.physics.ox.ac.uk/~yi/yyiso.html\\
http://www.astro.yale.edu/demarque/yyiso.html\


\section{Comparisons with intermediate-age star clusters}

Identifiable 
gaps in the stellar distribution, which can be used to evaluate  
the extent of core overshoot, are best observed in the CMDs 
of intermediate age 
clusters, which are generally more populous than the young 
star clusters. 
Such gaps have been observed 
in open clusters in
the Galactic disk, and more recently in intermediate age clusters of the
Large Magellanic Clouds.  
In this section, we compare the $Y^2$ OS isochrones to a few well observed 
open cluster CMDs to illustrate the improvement of $Y^2$ OS isochrones.
We emphasize that these comparisons merely show the compatibility of 
the OS isochrones with some of the available observational data, which 
cover only a small portion of parameter space.  Indeed the glitches  
encountered with the original $Y^2$ isochrone interpolations 
mostly occurred for compositions and ages for which we do not have 
observed CMDs in the Galaxy or the LMC, 
but which are important in the systematic 
modeling of stellar populations.   

\subsection{Open star clusters in the Galaxy}

Field star contamination is a problem 
in establishing open cluster membership
(Kozhurina-Platais et al. 1995; Platais et al. 2003).  In addition, the 
presence of binary 
stars modifies the stellar distribution near the main sequence turnoff. 
We present CMD fits for selected  
Galactic open clusters for which proper motion membership 
and binary membership 
have been established by the WOCS (WIYN open cluster studies) collaboration 
(Mathieu 2000; Sarajedini, Mathieu \& Platais 2003).  We have 
selected the following WOCS clusters: NGC 3680 (Khozurina-Platais 
et al. 1997), NGC 2420 (Demarque et al. 1994),
M67 (Girard et al. 1989; Sandquist 2004),
and NGC 6791 (Kaluzny \& Udalski 1992), in order of increasing age. 
NGC 3680 and M67 are close to solar in [Fe/H]. NGC 6791 
is more metal rich than the Sun (Peterson \& Green 1998), 
while the thick disk cluster NGC 2420 is   
metal poor (Grocholski \& Sarajedini 2003).

In order to fit isochrones to observed CMDs of star clusters, 
we first chose [Fe/H] for each cluster and determined the best-fitting
age by adjusting the cluster distance modulus and reddening.  The adopted 
parameters for each cluster are listed in Table~2.  
Although it is not straightforward to compare our adopted values of the
distance modulus and reddening with literature values,
due to the uncertainties of metallicity and metallicity dependence 
of the models, our values agree in general with literature values, e.g.,
$(m-M)_{V}$=10.20 and E(B-V)$_{o}$=0.06 of NGC 3680 (Anthony-Twarog \& Twrog 2004),
$(m-M)_{V}$=12.10 and E(B-V)$_{o}$=0.04 of NGC 2420 (von Hippel \& Gilmore 2000),
$(m-M)_{V}$=9.65 and E(B-V)$_{o}$=0.038 of M67 (Sandage et al. 2003),
and $(m-M)_{V}$=13.42 and E(B-V)$_{o}$=0.17 of NGC 6791 (Chaboyer et al. 1999a; Kaluzny \& Rucinski 1995).
Within the 
uncertainties, the ages are in 
each case within the range of recent age determinations based on 
theoretical isochrones.  
Since the total metallicity Z includes both [Fe/H] and [$\alpha$/Fe], 
we determined ages for each 
[Fe/H] for several values of the 
$\alpha$ enrichment parameter. 

Figure~3 shows three isochrones, depending on the $\alpha$ parameter,
fitted to the observed CMD for each cluster.
Within the uncertainties which are still very large (see e.g. the recent 
compilation by Salaris et al. 2004), 
the ages listed in Table~2 are compatible with 
the ages derived in other investigations using independent 
isochrone calculations. For NGC 3680, we 
assign an age in the range 1.3--1.7 Gyr, to be compared with 1.3--1.7 Gyr 
from Kozhurina-Platais et al. (1997). For NGC 2420, we find 1.4--2.0 Gyr, 
using [Fe/H]=$-$0.27, which is compatible with the age  
of $2.4\pm0.2$ Gyr previously derived by   
Demarque et al. (1994) using [Fe/H]=$-$0.7 (and more recently by 
Grocholski \& Sarajedini 2003). For M67, 3.6 Gyr (for solar 
abundance) is in good agreement with the recently derived age of 3.7 Gyr, 
based on a model including detailed diffusion effects (Michaud et al. 2004).   
Finally, in the case of NGC 6791, the
only good fit was found for the solar mixture.
Our age estimate for NGC 6791 is 8.0 Gyr, to be compared to 
$8.0\pm0.5$ Gyr derived by Chaboyer et al. (1999a).

It is clear from inspection of Figure~1 and Figure~2 that as satisfory 
fits as in Figure~3 could not be achieved  with the old-version 
isochrones in the transition age region. 
We must emphasize, however, that the original approach was tailored to 
fitting isochrones to CMD's specifically for the ages and compositions of 
nearby stars clusters in the Galaxy.  The need to improve the ramping 
and interpolation 
procedure became more evident when the $Y^2$ isochrones became increasingly 
utilized for more systematic studies of population synthesis 
covering more complete samplings of the parameter space.   

\subsection{Three clusters in the Large Magellanic Cloud}
The CMDs of intermediate age open clusters in the Large Magellanic 
Cloud (LMC) are of particular interest because they are 
more metal poor than in the Galaxy.
Because of their large distances from us, 
it is not possible to test membership through proper motion 
studies. Few radial velocity measurements are available.  
Although the membership problem is not as severe 
in the LMC as in the Galaxy because the clusters are more populated 
and field star contamination is less important,  
estimates of convective core overshoot
from their CMDs cannot be made as securely as in Galactic open clusters.
In addition, the binary star population is significant in these 
clusters (Woo et al. 2003). 
Figure~4 shows a first-cut comparison (ignoring binary stars) of 
our $Y^2$ OS isochrones to three CMDs
of LMC clusters obtained with the Very Large Telescope at the European Southern
Observatory. 
Isochrone fits to the 
CMDs for NGC 2173, SL 556 and NGC 2155 are displayed,  assuming 
the metallicity Z = 0.004 (Gallart et al. 2003). More detailed analyses 
of the three 
CMDs, which provide independent age estimates  based on the Yale and Padova
isochrones, have been published by Woo et al. (2003) and 
Bertelli et al. (2003), respectively.  
Once [Fe/H] is fixed, it is not possible to use the same distance modulus 
and reddening for each different value of $\alpha$. 
Since the purpose of the CMD fitting is to demonstrate the 
improved fit near the turnoff of $Y^2$ OS 
isochrones, rather than to determine exact cluster parameters,
we used a fixed metallicity Z, instead of a fixed value of [Fe/H],
and the same age for each $\alpha$ parameter.
As in the case of Galactic open clusters, the new isochrones show 
a good fit to the observed CMDs.

\section{Future prospects: observations and modeling}

On the observational side, 
two space asteroseismology missions, WIRE (Buzasi 2000) and MOST 
(Matthews et al. 2000) have begun to provide data on a few 
selected bright star targets.  The analysis of these data 
holds great promise for testing the 
internal structure of these stars and the chemical composition profile 
in their interiors (Basu et al. 2004; Guenther \& Brown 2004).
In particular, seismology is expected to   
reveal the presence and extent of convective cores and 
envelopes in stellar interiors (Audard et al. 1995; 
Chaboyer et al. 1999b).

On the side of theoretical convection modeling, 
it is expected that future progress in core overshoot 
understanding will result from  
a combination of analytical and numerical hydrodynamical 
approaches. As an example, the implementation of 
the Kuhfu\ss~ theory in the YREC stellar evolution code has 
already yielded a more  
physically realistic description of the overshoot process 
(Straka et al. 2004). Because core overshoot reaches its  
full development in a narrow mass interval 
above $M^{conv}_{crit}$,  
a fine mass grid of evolutionary tracks, i.e. 
with a mass increment as small as 0.001 $M_\odot$, will be 
required for future isochrone construction in the transition region.  
Already, extremely fine mass grids have been found to be necessary to 
fully exploit the information contained in 
high quality asteroseismic data (Guenther \& Brown 2004).
 The confrontation of these models with seismic observations from 
the WIRE and MOST space missions will in the next few years 
provide the foundation for a physical theory of convective core 
overshoot on which to base the next generation of isochrones. 

\acknowledgments
We are indebted to Ata Sarajedini for kindly sending us cluster CMD data.
This research has been supported in part by 
grants NAG5-8406 and NAG5-13299 from NASA to Yale University (PD)
and by PPARC Theoretical Cosmology Rolling Grant PPA/G/O/2001/00016 
(SKY).     YCK 
is supported by the Astrophysical Center for the Structure and Evolution 
of the Cosmos (ARCSEC) of Korea Science and Engineering Foundation 
(KOSEF) through the Science Research Center (SRC) Program.\\

\clearpage

\begin{table}
\label{table1}
\caption{Adopted values of $M^{conv}_{crit}$} 
\begin{tabular}{lccc} \hline
       
Composition &   &   &   \\ \hline
~~~~~~~~~[$\alpha$/Fe] &0.0 &0.3 &0.6\\  \cline{2-4}
x76997z00001 & 2.1 & 2.1 & 1.9\\
x7697z0001 & 1.6 & 1.6 & 1.5\\
x7688z0004 & 1.4 & 1.5 & 1.3\\
x767z001 & 1.3 & 1.4 & 1.2\\
x758z004 & 1.2 & 1.3 & 1.2\\
x749z007 & 1.2 & 1.2 & 1.2\\
x74z01 & 1.2 & 1.2 & 1.2\\
x71z02 & 1.2 & 1.2 & 1.2\\
x65z04 & 1.1 & 1.1 & 1.0\\
x59z06 & 1.1 & 1.1 & 1.0\\
x53z08 & 1.0 & 1.0 & 0.9\\ 
\end{tabular}
\end{table}

\begin{table}
\label{table2}
\caption{Selected parameters for each cluster} 
\begin{tabular}{lcrrrrr} \hline
       
Name         & [Fe/H] & $(m-M_{V})$  & E(B-V) &           &age          &             \\ 
             &        &             &        & [$\alpha$/Fe]~   0.0 & 0.3 & 0.6 \\                 
NGC 3680     &  0.0   & 10.4        &0.04    &               1.7 & 1.4 & 1.2 \\
NGC 2420     & -0.27  & 12.2        &0.05    &               2.0 & 1.7 & 1.4 \\
M67          &  0.0   & 9.9         &0.03    &               3.6 & 3.4 & 2.8 \\
NGC 6791     &  0.3   & 13.5        &0.17    &               8.0 & -   & -   \\
\end{tabular}
\end{table}

\clearpage

\begin{figure}
\epsscale{0.4}
\plotone{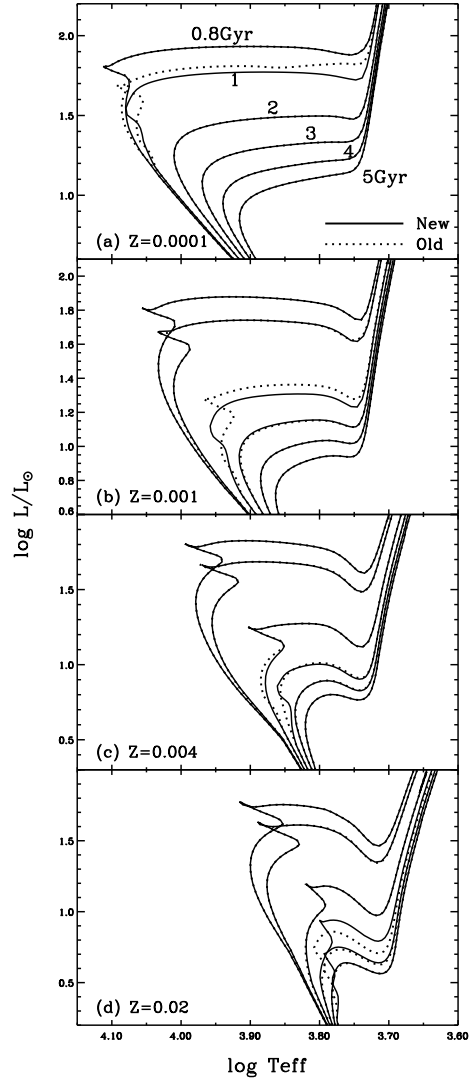}
\caption{Comparison between old ($dotted$) and revised ($continuous$)
isochrones in the convective core transition region.  Note that the 
transition occurs at a different age range for different chemical 
composition. The change due to the improvement in the overshoot
treatment in this study is only visible near the ``transition age''.}
\end{figure}

\begin{figure}
\epsscale{1}
\plotone{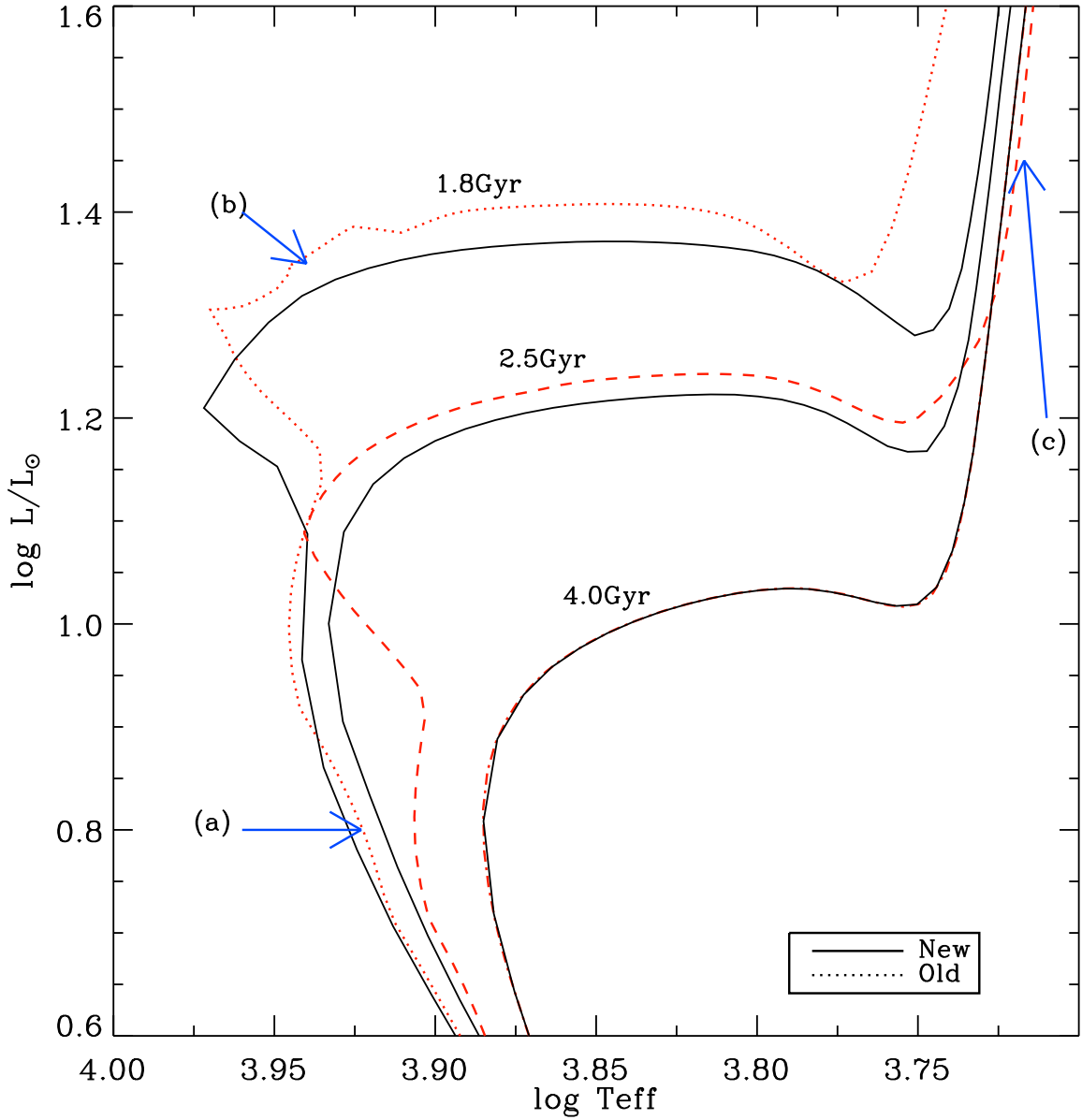}
\caption{The problems with our previous isochrones due to their
delta-function prescription for the overshoot are illustrated as
(a), (b), and (c). See the text for details. 
The problems, apparent for the two interpolated
isochrones (1.8 and 2.5~Gyr) near the transition age (1.9--2.0~Gyr), are all
resolved in the new isochrones via our overshoot-ramping prescription.
}
\end{figure}

\begin{figure}
\plotone{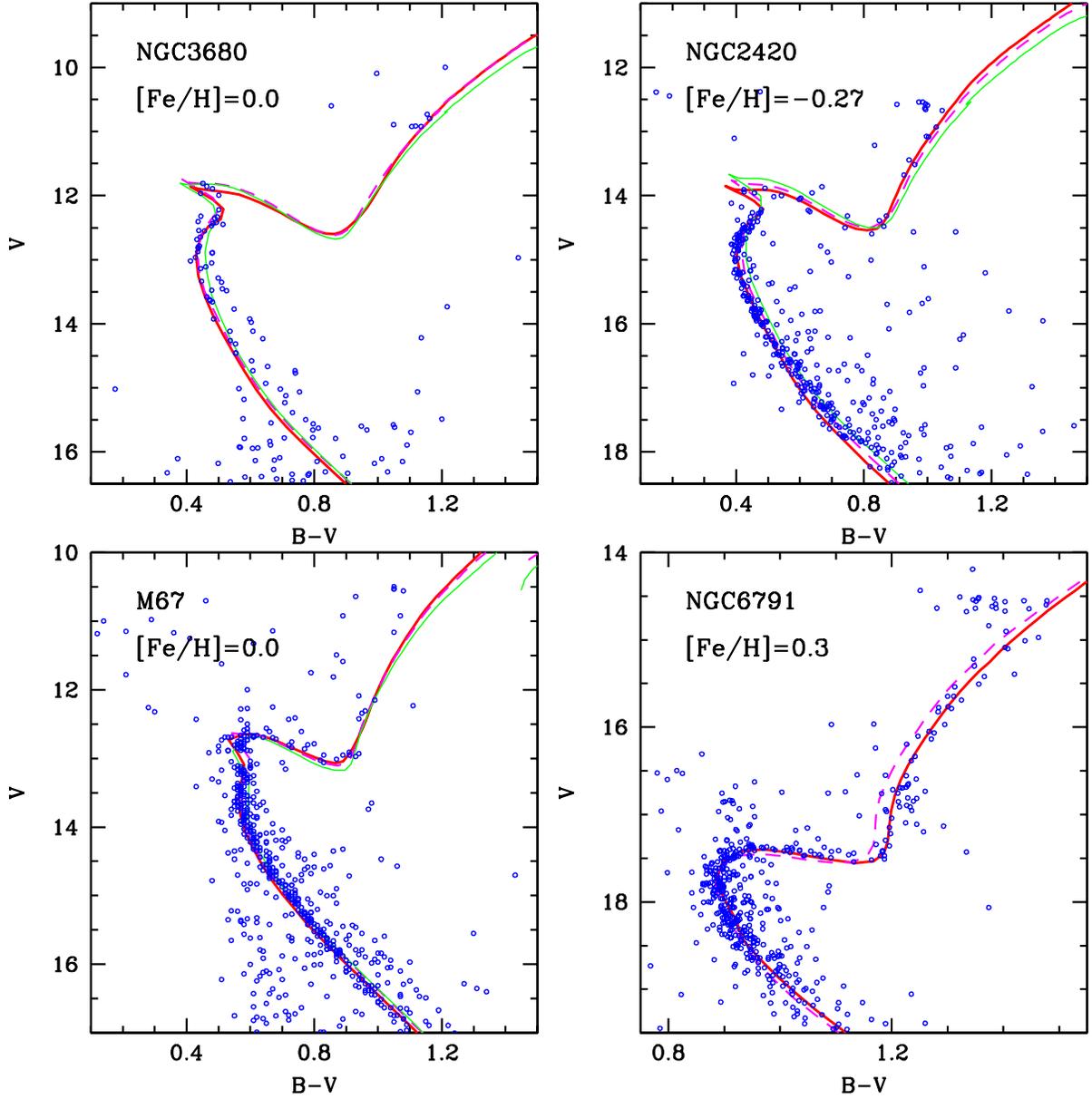}
\caption{The best-fitting isochrones with observed CMDs of four Galactic 
open clusters.
{\it Thick line:} $Y^2$ OS isochrone with [$\alpha$/Fe]=0.0, 
{\it dashed line:} [$\alpha$/Fe]=0.3, 
{\it thin line:} [$\alpha$/Fe]=0.6.
The smoothed turn-off shape of the new isochrones represents the 
observed CMD very well.
Note that the isochrone ages for the same [Fe/H] are different due to 
different Z values.
For NGC 6791, we could not fit the CMD with [$\alpha$/Fe]=0.6 isochrone. 
Even with the [$\alpha$/Fe]=0.3
isochrone, the fit is not acceptable, especially on the giant branch.
}
\end{figure}

\begin{figure}
\plotone{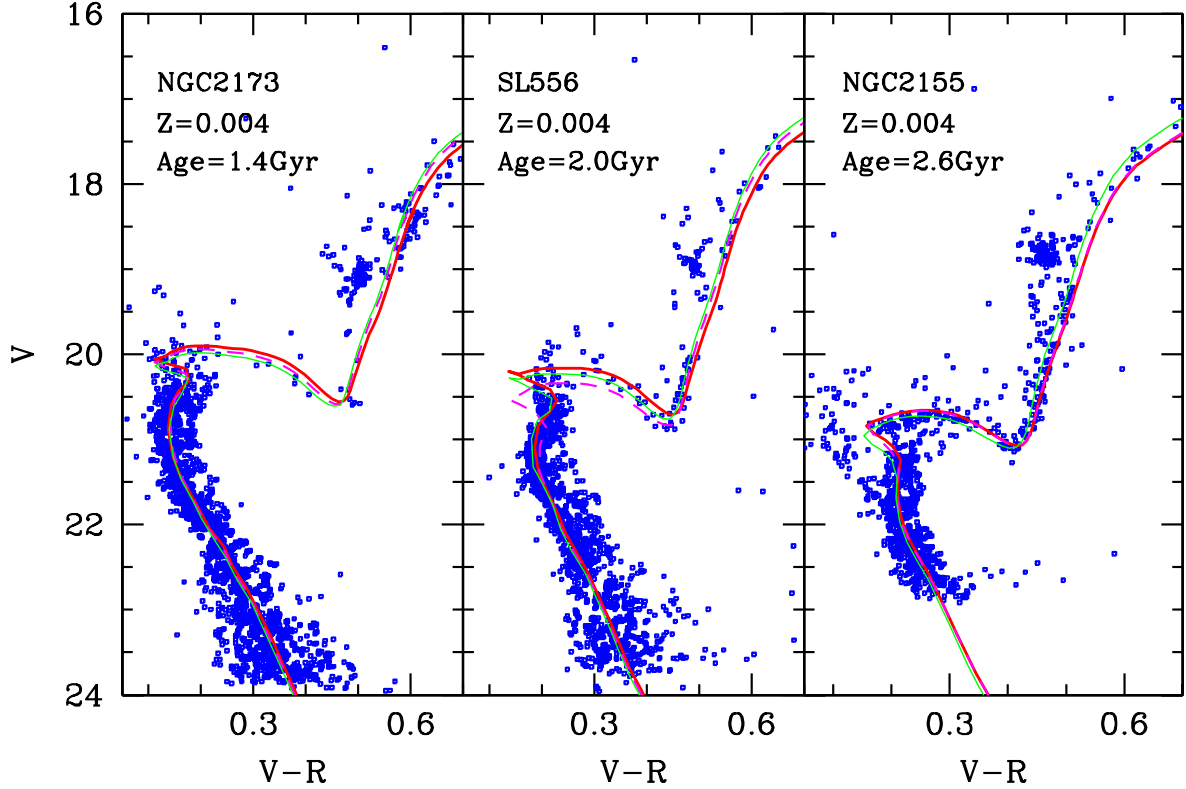}
\caption{The best-fitting isochrones with observed CMDs of three LMC clusters.
{\it Thick line:} $Y^2$ OS isochrone with [$\alpha$/Fe]=0.0, 
{\it dashed line:} [$\alpha$/Fe]=0.3, 
{\it thin line:} [$\alpha$/Fe]=0.6.
Isochrones are compared with the part of CMDs where the single 
stars are located. In constrast to the Galactic open clusters,
these LCM clusters are well populated and many unresolved binary 
stars are present. (For details on the effect of binaries  
on the synthetic CMD fitting, see
Woo et al. 2003). The new isochrones represent the turn-off 
region of the
observed CMD very well. For each cluster, 
($m-M_{V})=18.9,18.6,18.6$ and E(V-R)=0.08,0.06,0.03 are
used for NGC 2173, SL 556 and NGC 2155 respectively. 
}
\end{figure}
\end{document}